\documentclass[12pt]{article}

\usepackage{amssymb,amsmath}

\usepackage{graphicx}

\newcommand{\be}{\begin{equation}}
\newcommand{\ee}{\end{equation}}
\newcommand{\Dlt}{\Delta}
\newcommand{\dlt}{\delta}

\newcommand{\bbr}{{\bf r}}

\newcommand{\bn}{{\bf n}}
\newcommand{\bbe}{{\bf e}}
\newcommand{\bM}{{\bf M}}
\newcommand{\bH}{{\bf H}}
\newcommand{\bR}{{\bf R}}
\newcommand{\bS}{{\bf S}}
\newcommand{\bB}{{\bf B}}
\newcommand{\bt}{\beta}

\newcommand{\al}{\alpha}

\newcommand{\gm}{\gamma}
\newcommand{\om}{\omega}

\newcommand{\lbd}{\lambda}

\newcommand{\rgl}{\rangle}
\newcommand{\lgl}{\langle}

\begin{document}

\begin{center}

{\Large{\bf
Coherent spin dynamics of nanomolecules and magnetic nanoclusters} \\ [5mm]

V.I. Yukalov$^{1}$ and E.P. Yukalova$^{2}$} \\ [3mm]

{\it $^1$Bogolubov Laboratory of Theoretical Physics, \\
Joint Institute for Nuclear Research, Dubna 141980, Russia \\ [2mm]

$^2$Laboratory of Information Technologies,  \\
Joint Institute for Nuclear Research, Dubna 141980, Russia}

\end{center}

\vskip 3cm

\begin{abstract}
Spin dynamics of nanomolecules and nanoclusters are analyzed. The nanosizes
of these objects make it possible to consider them as single-domain magnets
with a large total spin, where the motion of the spins of all atoms, composing 
a nanocluster, occurs in a coherent way. Another meaning of coherence in spin
dynamics is the coherent spin motion of several nanomolecules or nanoclusters.
Different approaches for treating spin dynamics are compared and the main 
mechanisms influencing the spin motion are studied. Spin dynamics of separate
magnetic nanomolecules and nanoclusters are investigated, as well as the spin
dynamics of the ensembles of these nano-objects.      
\end{abstract}

\section{Introduction}

Magnetic nanomolecules and nanoclusters enjoy many similar properties because
of which the dynamics of their magnetization can be described by the same type 
of equations. This is why, we consider both these nano-objects together. Of 
course, there is difference in their structure and parameters which we shall 
take into account and characterize them by the appropriate models. The detailed
description of general physical properties and applications of different 
magnetic nanoparticles can be found in review articles [1-9]. Here we briefly 
mention those of the properties and parameters that will be necessary for the 
following consideration.

It is worth stressing that there exist two types of magnetic nanoparticles.
One large class consists of nanoclusters and nanomolecules, whose magnetic 
moments are formed by electron spins. Another type includes nanomolecules
that possess magnetic moments solely due to polarized proton spins. Examples 
are propanediol C$_3$H$_8$O$_2$, butanol C$_4$H$_9$OH, and ammonia NH$_3$. 
In such nanomolecules, there is no any other magnetic moment except that caused 
by polarized protons. So, here the proton magnetic moment is not a contribution, 
but the main object. 

The magnetic moment of an atom is composed of electron and proton moments, 
with the electron magnetic moment 
$\mu_e = - g_e \mu_B S = \hbar \gamma_e S = - \mu_B$ and the proton magnetic 
moment $\mu_p = g_p \mu_N S = \hbar \gamma_p S$, where $g_e = 2$  and 
$g_p = 5.586$ are the electron and proton Land\'{e} factors, 
$\mu_B = |e| \hbar / 2 m_e$  and $\mu_N = |e| \hbar / 2 m_p$ are the Bohr and
nuclear magnetons, $\gamma_e$ and $\gamma_p$ are the electron and proton 
gyromagnetic ratios. Since the proton mass is larger than that of an electron,
$m_p / m_e \sim 10^3$, the proton magnetic moment is essentially smaller,
$\mu_p / \mu_e \sim 10^{-3}$. The electron and proton radii are 
$r_e \sim 10^{-15}$cm and $r_p \sim 10^{-13}$cm, respectively. An atom is 
called magnetic, when its total magnetic moment is nonzero. The total spin of 
a magnetic atom can be between 1/2 and $S \sim 10$, hence, its magnetic moment 
can be of order $1 \mu_B - 10 \mu_B$. Atom radii are of order 
$r_A \sim 10^{-9} - 10^{-8}$cm. Examples of magnetic atoms are Fe (Iron),
Co (Cobalt), Ni (Nickel), Gd (Gadolinium), and Cr (Chromium). 

Magnetic nanomolecules are composed of many magnetic atoms an, as is clear 
from their name, are of the nanometer size. An important property of a 
magnetic nanomolecule is that its total magnetic moment can be treated as 
being due to an effective total spin. Generally, the molecule spin can be 
directed either up or down, with an energy barrier between these directions 
of order $E_A \sim 10 - 100$ K. At high temperatures, above the 
{\it blocking temperature} $T_B \sim 1 - 10$ K, a magnetic molecule behaves 
as a superparamagnetic particle, whose spin randomly oscillates between the 
up and down positions. While below the blocking temperature the spin is 
frozen in one of the directions. 

Magnetic nanoclusters are also made of magnetic atoms that are assembled 
together in a random way. This distinguishes them from magnetic molecules, 
where atoms are strictly connected by chemical bonds. The sizes of nanoclusters 
can be in the range between 1 nm and 100 nm, containing about $100 -10^5$ 
atoms. These values define the {\it coherence radius} $R_{coh}$, below which 
a nanocluster is in a single-domain state and can be treated as a large particle 
with an effective spin. A cluster, with a size larger than $R_{coh}$, separates 
into domains with opposite magnetizations. Similarly to magnetic molecules at 
low temperature, the magnetic moment of a nanocluster, below the blocking 
temperature $T_B \sim 10 - 100$ K, is frozen in one of two possible 
directions. The effective spin of a nanocluster is formed by electron spins
and can be as large as $S \sim 100 - 10^5$.  

The often considered nanoclusters are made of the magnetic atoms of Fe, Ni, 
and Co. They can be made of oxides, such as NiO, Fe$_2$O$_3$, NiFe$_2$O$_4$ 
or alloys, such as Nd$_2$Fe$_{14}$B, Pr$_2$Fe$_{14}$B, Tb$_2$Fe$_{14}$B, 
DyFe$_{14}$B, Pr$_2$Co$_{14}$B, Sm$_1$Fe$_{11}$Ti$_1$, Sm$_1$Fe$_{10}$V$_2$, 
Sm$_2$Fe$_{17}$N$_{23}$, Sm$_2$Fe$_{17}$C$_22$, Sm$_2$Co$_{17}$, Sm$_2$Co$_5$. 
To protect nanoclusters from oxidation, one coat them with graphene or noble 
metals, forming the double-component nanoclusters, such as Fe-Au, Co-Au , 
Co-Ag, Co-Cu, Co-Pt, Co-Pd, Ni-Au, Ni-Ag, Ni-Pd, and Mn$^-$-Au. The coating
is done be means of chemical reactions or laser ablation techniques. The 
nanoclusters are produced by employing thermal decomposition, microemulsion
reactions, and thermal spraying.  

Magnetic nanoclusters and nanomolecules find numerous applications, among 
which we can mention magnetic chemistry, biomedical imaging, medical treatment, 
genetic engineering, waste cleaning, information storage, quantum computing, 
and creation of radiation devices. Since both nanomolecules and nanoclusters
possess many common properties and can be considered as single particles with 
a large spin, we shall often talk on nanoclusters, implying that similar 
effects can be realized with both of them, molecules as well as clusters.    

The use of these nano-objects requires the existence of two properties that 
contradict each other. From one side, to be able to keep memory, a cluster 
has to enjoy a stable state with its spin frozen in one direction. But from 
another side, in order to be able to manipulate the cluster magnetization, 
there should exist a way of suppressing the anisotropy. And it is necessary 
that the spin manipulation could be done sufficiently fast, so that the 
cluster magnetization could be quickly reversed. Recall that thermal 
reversal is characterized by the Arrhenius law giving the longitudinal 
relaxation time $T_1 \sim \exp\{E_A/k_BT\}$, where $E_A$ is the anisotropy 
energy, so that, at temperatures below the blocking temperature, the 
magnetization is frozen.      

Magnetization reversal can be realized by different methods, by applying
transverse constant or alternating magnetic fields and short magnetic field 
pulses [10]. To achieve fast reversal, one needs to find optimal values for 
the amplitude, frequency, and duration of such field pulses. 

A very efficient method of achieving ultrafast magnetization reversal of 
magnetic nanoclusters has been suggested [11] by employing the acceleration
effect caused by a resonator feedback field. The efficiency of this method
is due to self-optimization of the spin motion producing the resonator 
field acting back on the spins. Historically, this effect was described by
Purcell [12] and considered by Bloembergen and Pound [13] using classical 
phenomenological equations. Such equations are not sufficient for describing 
different regimes of spin motion. Microscopic theory of spin dynamics has 
been developed being applied to polarized proton spins of such molecules as 
propanediol C$_3$H$_8$O$_2$, butanol C$_4$H$_9$OH, and ammonia NH$_3$ 
(see review articles [4,14]) and to magnetic molecules [15-19].

The aim of the present paper is threefold. First, we concentrate on the spin 
dynamics of nanoclusters, comparing the peculiarity of their spin motion with
that of proton and molecular spins. Second, we analyze the role of other 
effects, such as the Nyquist-noise triggering and Dicke correlation, studying
their influence on the spin dynamics of nanoclusters. We show that these 
effects are negligible as compared to the Purcell effect. And, third, we 
compare different approaches to describing spin dynamics, demonstrating the
advantage of using a microscopic approach based on quantum equations of motion.

\section{Phenomenological classical equations}

Dynamics of the magnetic moment ${\bf M}$ of a magnetic particle is usually 
described by the classical equation
\be
\label{1} 
 \frac{d\bM}{dt} = - |\gm_S| \bM \times \bH_{eff} + \bR \;  ,
\ee
in which $\gamma_S$ is the giromagnetic ratio of the particle with spin $S$
and $\bf R$ is a relaxation term. The effective magnetic field is given by 
the variational derivative $\bH_{eff} = - \; \dlt E / \dlt\bM$ of the particle 
energy $E$. The length of the magnetic moment is conserved, when the right-hand 
side of the equation $d\bM^2 / dt = 2 \bM \cdot \bR$ is zero. 

Choosing the relaxation term in the form
\be
\label{2}
 \bR = - \; \frac{\al|\gm_S|}{M} \; \bM \times 
\left ( \bM \times \bH_{eff} \right ) \;  ,
\ee
one gets the Landau-Lifshitz equation, where $\alpha$ is a dissipation 
parameter and $M \equiv |{\bf M}|$. Under form (2), $|{\bf M}|$ is conserved. 
The equation was initially derived [20] for describing energy dissipation 
in the process of magnetic domain wall motion inside bulk ferromagnetic matter. 
Though it is often applied for treating the dynamics of ferromagnetic 
particles [21].
 
Taking the relaxation term as
\be
\label{3}
\bR = \frac{\al}{M} \; \bM \times  \frac{d\bM}{dt} \;  ,
\ee
one comes to the Gilbert equation [22]. This equation, up to a renotation 
of parameters, is equivalent to the Landau-Lifshitz equation. Hence, it has
the same region of applicability, though it is also used for describing the
magnetization rotation of magnetic particles [10].   

Another form of the relaxation term has been advanced by Bloch [23] as
\be
\label{4}
 \bR = - \; \frac{M_x-M_x^*}{T_2}\; \bbe_x \; - \;
\frac{M_y-M_y^*}{T_2}\; \bbe_y \; - \; 
\frac{M_z-M_z^*}{T_1}\; \bbe_z \; ,
\ee
where ${\bf M}^*$ is an equilibrium magnetization, ${\bf e}_\alpha$ are 
unit coordinate vectors, and the relaxation parameters are characterized 
by the longitudinal relaxation time $T_1$ and transverse relaxation time 
$T_2$. The latter is also called the dephasing time. For an ensemble of $N$ 
magnetic particles with a large average spin polarization
\be
\label{5}
 s \equiv \frac{1}{SN} \; \sum_{j=1}^N \; \lgl S_j^z \rgl \;  ,
\ee
the transverse term has to be renormalized [16,24] as 
$1 / \widetilde T_2 = (1-s^2) / T_2$ .

The Landau-Lifshits equation has a single dissipation parameter $\alpha$ 
and preserves spherical symmetry, thus, describing isotropic magnetization 
rotation. Because of these properties, it is appropriate for bulk macroscopic 
ferromagnetic matter with spherical magnetic symmetry. It may also be used
for magnetic clusters, possessing this symmetry, which, however, is a rather 
rare case. 

The Bloch equation has two relaxation parameters, $T_1$ and $T_2$. Therefore
it can describe more general situation of anisotropic relaxation, which is 
more realistic for treating nanoclusters in a medium or below the blocking 
temperature, when $T_2 \ll T_1$. The Bloch equations have been employed for 
considering the electron and nuclear spin motion in a strongly coherent 
regime [25-29] and for spin-polarized $^{129}$Xe gas [30]. But these equations
cannot describe the whole process of spin relaxation starting from an 
incoherent quantum stage, for which a microscopic approach is necessary 
[29,31,32]. The initial stage of spin relaxation is triggered by quantum spin 
fluctuations that can be identified with nonequilibrium spin waves [4,16,31-33].

\section{Microscopic quantum approach}

In a self-consistent quantum approach, we start with a microscopic spin 
Hamiltonian $\hat{H}$ that is a functional of spin operators ${\bf S}$.
The evolution equations are given by the Heisenberg equations of motion 
\be
\label{6}
 i\hbar \; \frac{d\bS}{dt} = [ \bS , \; \hat H ] \; .
\ee
The advantage of using the quantum approach is in the following. First, it
takes into account quantum effects that can be important for small clusters.
Hence, it is more general. Second, at the initial stage of free spin 
relaxation, quantum spin fluctuations are of principal importance, being the 
triggering mechanism for starting the spin motion. Third, being based on an
explicit spin Hamiltonian makes it possible to control the used approximations
and to have well defined system parameters.

We assume that a magnetic cluster is inserted into an magnetic coil, of $n$ 
turns and length $l$, of a resonant electric circuit characterized by 
resistance $R$, inductance $L$, and capacity $C$. The coil axis is taken 
along the axis $x$. Moving magnetic moments induce in the coil the electric 
current $j$ described by the Kirchhoff equation
\be
\label{7}
L \; \frac{dj}{dt} + Rj + \frac{1}{C} \int_0^t j \; dt = - \; 
\frac{d\Phi}{dt} + E_f \;   , 
\ee
in which the magnetic flux $\Phi = 4\pi n M_x / cl$ is formed by the mean 
transverse magnetization $M_x = \mu_0 \sum_{j=1}^N \; \lgl S_j^x \rgl$, where
$\mu_0 \equiv \hbar\gm_S$. Here $E_f$ is an additional electromotive force, 
if any. The resonator natural frequency and circuit damping, respectively, are
\be
\label{8}
 \om = \frac{1}{\sqrt{LC} } \; , \qquad \gm = \frac{R}{2L} \;  .
\ee 
The coil current creates the magnetic field
\be
\label{9}
 H = \frac{4\pi n}{cl}\; j  
\ee
that is the solution to the equation
\be
\label{10}
 \frac{dH}{dt} + 2\gm H + \om^2 \int_0^t H(t') \; dt' = -
4\pi\eta \; \frac{dm_x}{dt} \;  ,
\ee
where $\eta \equiv V/V_{coil}$ is the filling factor and
$$
m_x \equiv \frac{M_x}{V} = \frac{\mu_0}{V} \; 
\sum_{j=1}^N \; \lgl S_j^x \rgl   
$$
is the transverse magnetization density. The external electromotive force
is omitted. The field $H$ is the feedback field, created by moving spins 
and acting back on them.

\section{Dynamics of a single nanocluster}

The typical Hamiltonian of a nanocluster is
\be
\label{11}
  \hat H = - \mu_0 \bB \cdot \bS - D (S^z)^2 + D_2 (S^x)^2 +
D_4 \left [ (S^x)^2 (S^y)^2 + (S^y)^2 (S^z )^2 + (S^z)^2 (S^x)^2
\right ] \; ,
\ee
where the total magnetic field
\be
\label{12}
\bB = B_0 \bbe_z + B_1 \bbe_x + H \bbe_x
\ee
consists of an external constant field $B_0$, weak transverse anisotropy 
field $B_1$, and the feedback resonator field $H$. The anisotropy parameters 
$D, D_2, D_4$ are defined by the particular type of considered nanoclusters. 

The main attention will be payed to the investigation of spin dynamics 
starting from a strongly nonequilibrium initial state, where the magnetization 
is directed opposite to the constant external magnetic field $B_0$.  
 
First, we study the influence of the thermal Nyquist noise of the coil in 
order to understand whether it can trigger the spin motion in a nanocluster.
For the thermal-noise relaxation time, we find
\be
\label{13}
 t_T = \frac{4\gm V_{coil} }{\hbar\gm_S^2\om} \; \tanh \left (
\frac{\om}{2\om_T} \right ) \;  ,
\ee
where $\om_T \equiv k_BT / \hbar$ is the thermal frequency defined by 
temperature $T$. At low temperatures, below the blocking temperature, say 
at $T = 1$ K, we have $\omega_T \sim 10^{12}$ s$^{-1}$. Then the 
thermal-noise relaxation time is
\be
\label{14}
 t_T \simeq \frac{2\gm V_{coil} }{\hbar\gm_S^2\om_T} \qquad
\left ( \frac{\om}{\om_T} \ll 1 \right ) \; .
\ee
On the other side, for the reversal time, caused by the resonator feedback 
field, we have 
\be
\label{15}
t_{rev} \simeq \frac{V_{coil} }{\pi\hbar\gm_S^2 S} \;   .
\ee
The ratio of the latter to the thermal time (13) is 
$t_{rev} / t_T \sim \om_T / 2\pi\gm S$. For the typical values $T = 1$ K, 
$\gamma \sim 10^{10}$ s$^{-1}$, and $S \sim 10^3$, this ratio is small: 
$t_{rev} / t_T \sim 10^{-2}$. Therefore the thermal Nyquist noise does not 
play any role in the spin dynamics of a nanocluster.

We have accomplished numerical solution of the evolution equations for 
nanocluster parameters typical of Fe, Ni, and Co nanoclusters. The Zeeman 
frequency is taken as $\om_0 \equiv 2\mu_B B_0 / \hbar \sim 10^{11}$ s$^{-1}$.
For the feedback rate, we have 
$\gm_0 \equiv \pi\eta\hbar\gm_S^2 S / V_{coil} \sim 10^{10}$ s$^{-1}$.
The typical anisotropy parameters satisfy the relations
$D / (\hbar\gm_0) \sim 10^{-3}, D_2 / (\hbar\gm_0) \sim 10^{-3}, 
D_4 / (\hbar\gm_0) \sim 10^{-10}$. At the initial time, the spin is assumed 
to be directed along the axis $z$. The resonator natural frequency is taken 
to be in resonance with the Zeeman frequency defined by the field $B_0$. The 
behavior of the spin polarization (5) is shown in Fig. 1, where we compare 
the spin motion in the presence of the resonator ($h \neq 0$) and in the 
absence of the latter ($h = 0$). Clearly, without the resonator feedback field, 
the spin is blocked, while in the presence of the resonator, it reverses in 
short time $t_{rev} \sim 10^{-10}$ s.

\begin{figure}[h]
\includegraphics[width=14pc]{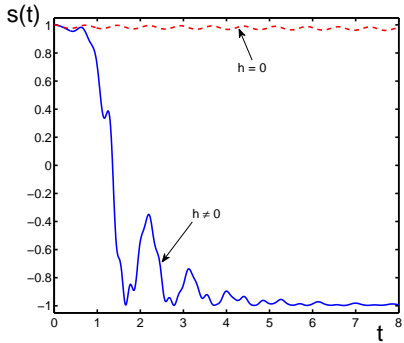}\hspace{0.5pc}%
\begin{minipage}[b]{21pc}
\caption{\label{Figure 1}Spin reversal of a single nanocluster, with parameters 
typical of nanoclusters made of Fe, Ni, and Co.}
\end{minipage}
\end{figure}

\section{Dynamics of nanocluster assemblies}

The ensemble of nanoclusters is described by the Hamiltonian
\be
\label{16}
 \hat H = \sum_i \hat H_i + 
\frac{1}{2} \; \sum_{i\neq j} \hat H_{ij} \;  ,
\ee
where the indices $i,j = 1,2,\ldots,N$ enumerate nanoclusters. The single
nanocluster Hamiltonians are
\be
\label{17}
 \hat H_i =  - \mu_0 \bB \cdot \bS_i -  D (S_i^z)^2 + D_2 (S_i^x)^2 +
D_4 \left [ (S_i^x)^2 (S_i^y)^2 + (S_i^y)^2 (S_i^z )^2 + (S_i^z)^2 (S_i^x)^2
\right ] \; ,
\ee
with the total magnetic field
\be
\label{18}
 \bB = B_0 \bbe_z + H \bbe_x \;  .
\ee
The interaction term takes into account the dipolar spin interactions
\be
\label{19}
  \hat H_{ij} = \sum_{\al\bt} D_{ij}^{\al\bt} S_i^\al S_j^\bt \; ,
\ee
through the dipolar tensor 
$D_{ij}^{\al\bt} = \mu_0^2 \left ( \dlt_{\al\bt} - 
3 n_{ij}^\al n_{ij}^\bt \right ) / r_{ij}^3$, in which
$r_{ij} \equiv |\bbr_{ij}|, \bn_{ij} \equiv \bbr_{ij} / r_{ij}$, and
$\bbr_{ij} \equiv \bbr_i - \bbr_j$.

One sometimes says that spin systems are similar to atomic systems, where 
transition dipoles are correlated by means of the photon exchange through
the common radiation field. This correlation leads to coherent atomic 
radiation called the Dicke superradiance [34]. One says that moving spins 
also radiate electromagnetic field that could yield the correlated spin
motion, in the same way as in the Dicke effect. To check whether this is 
so, we need to compare the time $t_{rad}$, required for inducing spin 
correlations through the common radiation field with the spin dephasing 
time $T_2$. As the radiation time [35,36] for nanoclusters, we have
\be
\label{20}
 t_{rad} = \frac{3c^3}{2\hbar\gm_S^2\om^3 S} \;  ,
\ee
while the spin dephasing time is
\be
\label{21}
 T_2 = \frac{1}{\hbar\rho\gm_S^2 S} \;  .
\ee
For the typical nanocluster density $\rho \sim 10^{20}$ cm$^{-3}$ and 
$S \sim 10^3$, the spin dephasing time is $T_2 \sim 10^{-10}$ s. While
for the radiation time (20), with $\omega \sim 10^{11}$ s$^{-1}$, we have
$t_{rad} \sim 10^8$ s $= 10$ years. The ratio of times (20) and (21) is 
extremely large: $t_{rad} / T_2 = 3c^3\rho / (2\om^3) \sim 10^{18}$.
This tells us that the spin motion in no way can be correlated through 
electromagnetic radiation. That is, the Dicke effect has no relation to 
the coherent spin motion. But spins can be correlated only through the 
Purcell effect requiring the presence of a feedback field caused by a 
resonator.

The feedback rate due to the resonator is
\be
\label{22}
\gm_0 = \pi \eta \rho \hbar \gm_S^2 S \;   .
\ee
The reversal time for $N$ correlated nanoclusters becomes
\be
\label{23}
 t_{rev} = \frac{1}{\gm_0} = \frac{V_{coil}}{\pi\hbar\gm_S^2 SN} =
\frac{t_{rev}^1}{N} \;  ,
\ee
where $t^1_{rev}$ is the relaxation time (15) for a single nanocluster
inside the same coil.

We solved the evolution equations for the nanocluster assemblies involving
the scale separation approach [4,14] that is a generalization of the 
Krylov-Bogolubov [37] averaging method. Four classes of spin objects have
been investigated. 

(i) Polarized nuclear materials, such as propanediol C$_3$H$_8$O$_2$, 
butanol C$_4$H$_9$OH, and ammonia NH$_3$, with the 
parameters: $S = 1/2, \rho = 10^{22} {\rm cm}^{-3}, T=0.1 {\rm K}, 
B_0 \sim 10^4 {\rm G}, \om_0 \sim 10^8 {\rm s}^{-1}, 
\lbd \sim 10^2 {\rm cm}, T_1 \sim 10^5 {\rm s}, T_2 \sim 10^{-5} {\rm s},
\tau \equiv 1/\gamma \sim 10^{-6} {\rm s}$ \; . Recall that in these 
nanomolecules the magnetization is due to polarized proton spins.

The following characteristic times are found: thermal-noise time
$t_T \sim 10^{16} {\rm s} \sim 10^9 \; {\rm years}$, radiation time
$t_{rad} \sim 10^{15} {\rm s} \sim 10^8 \; {\rm years}$, and reversal 
time $t_{rev} \sim 10^{-6} {\rm s}$.

Therefore, neither the Nyquist thermal noise nor the photon exchange 
through the radiated field play any role in the relaxation process. Spin
dynamics, resulting in the magnetization reversal, is completely due to 
the action of the resonator feedback field. As is explained above, the 
same concerns nanomolecules and nanoclusters 
 
(ii) Nuclear polarized ferromagnets, where proton spins are polarized and 
interact through hyperfine forces with electrons participating in forming
ferromagnetic order. In such materials, the electron subsystem plays the 
role of an additional resonator enhancing effective nuclear correlations. 
Being interested in the motion of nuclear spins, under a fixed mean electron 
magnetization, we find the reversal time $t_{rev} \sim 10^{-9}$ s.

(iii) Molecular magnets, such as Mn$_{12}$ and Fe$_8$, with the typical 
parameters: $S = 10, \rho = 10^{20} - 10^{21} {\rm cm}^{-3}, 
T_B = 1 {\rm K}, B_0 \sim 10^5 {\rm G}, \om_0 \sim 10^{13} {\rm s}^{-1},
\lbd \sim 10^{-2} {\rm cm}, 
\om_A \equiv E_A/\hbar \sim 10^{10} - 10^{12} {\rm s}^{-1}, 
T_1 \sim 10^5 - 10^7 {\rm s}, T_2 \sim 10^{-10} {\rm s}$.
The reversal time is $t_{rev} \sim 10^{-11}$ s.  
 
(iv) Magnetic nanoclusters composed of Fe, Ni, and Co, at $T = 1$ K, with 
the typical parameters: $S = 10^3, \rho = 10^{20} {\rm cm}^{-3},  
T_B = 10 - 40 {\rm K}, B_0 \sim 10^4 {\rm G} = 1 {\rm T}, 
\om_0 \sim 10^{11} {\rm s}^{-1}, \lbd \sim 1 {\rm cm},
T_1 \sim 10^{34} {\rm s} \sim 10^{27} {\rm years}, T_2 \sim 10^{-10} {\rm s}$. 
The reversal time can be very small reaching the value 
$t_{rev} \sim 10^{-12}$ s.

In the case of magnetic molecules and, especially, nanoclusters, because 
of their high spins, the system of many clusters can produce quite strong 
coherent radiation of the maximal intensity
\be
\label{24}
 I_{max} \sim \frac{2\mu_0^2}{3c^3} \; S^2 \om^4 N_{coh}^2 \;  ,
\ee
where $N_{coh} \sim \rho \lambda^3$ is the number of clusters in a coherent 
packet. The intensity of radiation of magnetic molecules, with 
$N_{coh} \sim 10^{14}$, is of order $I_{max} \sim 10^5 W$. And for magnetic 
nanoclusters, with $N_{coh} \sim 10^{20}$, the radiation intensity can reach 
$I_{max} \sim 10^{12} W$.

There can happen several regimes of spin dynamics depending on the initial 
spin polarization, the strength of a triggering pulse, and the effective 
coupling parameter
\be
\label{25}
 g \equiv \frac{\gm\gm_0\om_0}{\gm_2(\gm_2^2+\Dlt^2) } \; ,
\ee
where $\gamma_2 \equiv 1/T_2$ and $\Delta \equiv \omega - \omega_0$ is the 
detuning from resonance. These regimes for nanoclusters can be classified 
analogously to those occurring for nuclear magnets [4,16]: incoherent free 
relaxation, weakly coherent free induction, weakly coherent superradiance, 
strongly coherent pure superradiance, strongly coherent triggered superradiance, 
pulsing superradiance, and punctuated superradiance [38]. 

It is important to stress that the existence of magnetic anisotropy in 
magnetic nanoclusters does not preclude the realization of fast spin reversal,
provided the external magnetic field is sufficiently strong. The influence of
the anisotropy energy $E_A$ on the spin reversal of a nanocluster system 
is shown in Fig. 2, where $A \equiv E_A/\omega_0$. This regime corresponds to
pure spin superradiance.

\begin{figure}[h]
\includegraphics[width=14pc]{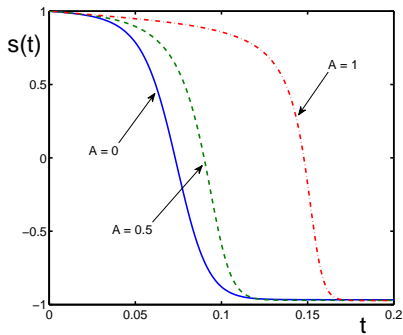}\hspace{0.5pc}%
\begin{minipage}[b]{21pc}
\caption{\label{Figure 2}Influence of magnetic anisotropy on the spin reversal 
of an ensemble of many nanoclusters, with the parameters typical of Fe, Ni, 
and Co.}
\end{minipage}
\end{figure}

Coherent dynamics in the spin assemblies, formed by magnetic nanomolecules, 
have an important difference from the spin dynamics in an ensemble of magnetic 
nanoclusters. Magnetic molecules are identical and form the systems with well
organized crystalline lattices. While magnetic nanoclusters vary in their shapes, 
sizes, and total spins, which results in an essential nonuniform broadening. 
Computer simulations, accomplished together with V.K. Henner and P.V. Kharebov,
demonstrate that this nonuniformity does not destroy coherent spin motion. A 
detailed analysis of the computer simulations, with nonuniform nanocluster 
distributions, will be presented in a separate publication.  
 
In conclusion, we have considered spin dynamics in magnetic nanomolecules and 
nanoclusters, starting from a strongly nonequilibrium state, with the 
magnetization directed opposite to the applied external magnetic field. We have
compared several methods of describing the spin dynamics, showing that a 
microscopic approach, based on the quantum equations of motion, is the most 
accurate. We also have analyzed the influence of different effects on spin 
dynamics. The effects of the Nyquist-noise triggering and of Dicke correlations 
are found to be negligible for spin systems. This principally distinguishes 
spin systems from atomic systems or quantum dot systems [39], where correlations, 
leading to coherent radiance, are caused by the Dicke effect of interactions 
through the common radiation field. 

The feedback field, developing in the resonator, reaches rather high values,
of the order of the applied constant magnetic field. Such a strong feedback 
field suppresses the influence of mutual cluster interactions. Generally, in 
an ensemble of nanoclusters of sufficiently high density, in addition to dipole 
interactions, there can appear exchange interactions [40] that can influence 
equilibrium properties of nanoclusters. But in the considered case of strongly
nonequilibrium spin dynamics, the exchange interactions are also suppressed by
the self-organized resonator feedback field.   

This important conclusion can be formulated as follows: {\it Coherent spin 
dynamics are completely governed by the Purcell effect that is caused by the 
action of the resonator feedback field}.

\end{document}